\documentstyle[12pt]{article}
\begin{document}

\small \hoffset = -1truecm \voffset = -2truecm
\title{\bf Rigorous proof of attractive nature for Casimir force of $p$-odd hypercube }
\author{
     Xin-zhou Li\footnote {e-mail address: kychz@shtu.edu.cn}\hspace{0.6cm}Xiang-hua Zhai\hspace{0.6cm}
  \\\footnotesize \it
    Department of physics,Shanghai Normal University, Shanghai 200234 ,China
      }
\date{}
\maketitle

\begin{abstract}
   The Casimir effect giving rise to an attractive force
between the configuration boundaries that confine the massless
scalar field is rigorously proven for odd dimensional hypercube
with the Dirichlet boundary conditions and different spacetime
dimensions $D$ by the Epstein zeta function regularization.
\end{abstract}

\vspace{1cm} \hspace{0.8cm} PACS number(s): 04.62.+v, 03.65.Ge
\newpage

In 1948 Casimir calculated an extraordinary property that two
uncharged metallic plates would have an attractive force in
vacuum[1]. This force is the strong function of $a$ and is
measurable only for $a<1\mu m$[2]. Boyer[3] numerically calculated
the Casimir force of a thin spherical shell, who found that the
sphere tends to be expanded. By using the mean approximation, the
stresses are directed outwards for a cubic cavity, which tend to
expand the cavity[4]. Therefore, one may imagine the spherical
shell to be deformed into a cubic shell, and expect that this
deformation does not change the resulting stresses. It is
interesting that the Casimir force of massless scalar field may
be repulsive for $p$-odd cavity with unequal edges[5]. On the
other hand, few physicists would nowadays argue against the
statement that the zeta function regularization procedure has
proven to be a very powerful and elegant technique. Rigorous
extension of the proof of Epstein zeta function regularization
has been obtained[6]. The generalized $\zeta$-function has many
interesting applications, e. g., in the piecewise string[7] and
branes[8]. In this paper, we rigorously proof that the Casimir
force is attractive for $p$-odd hypercube with the Dirichlet
boundary conditions and spacetime dimension $D$ less than
critical value $D_c$.

A Hermitian massless scalar field $\phi(t, x^{a}, x^{T})$ is
confined in the interior of  $(D-1)$-dimensional rectangular
cavity $\Omega$ with $p$ edges of finite lengths $L_{1},
L_{2},\cdots, L_{p}$ and $D-1-p$ edges with characteristic
lengths of order $\lambda\gg L_{a}$ where $i=1,\cdots,D-1;
a=1,\cdots,p; T=p+1,\cdots,D-1$. We consider the case of
Dirichlet boundary conditions,
i.e.,$\phi(t,x^a,x^T)|_{\partial\Omega}=0$. The Casimir energy
density is [5]

\begin{equation}
\varepsilon_p^D\left(L_{1}, L_{2},\cdots,
L_{p}\right)=-\frac{\pi^{(D-p)/2}}{2^{D-p+1}}\Gamma\left(-\frac{D-p}{2}\right)
E_p\left(\frac{1}{L_1^2},\frac{1}{L_2^2},\cdots, \frac{1}{L_p^2};
-\frac{D-p}{2}\right)\hspace{3cm}
\end{equation}

\noindent where the Epstein $\zeta$ function $E_p\left(a_1, a_2,
\cdots, a_p; s\right)$ is defined as

\begin{equation}
E_p\left(a_1, a_2, \cdots, a_p;
s\right)=\sum^{\infty}_{\{n\}=1}\left(\sum^{p}_{j=1}a_jn_j^2\right)^{-s}\hspace{3cm}
\end{equation}

\noindent Assuming $L_1=L_2=\cdots=L_p=L$, Eq.(1) can be reduced
to

\begin{equation}
\varepsilon_p^D(L)=\frac{L^{p-D}}{2^{D+1}}\sum^{p-1}_{q=0}(-1)^{q+1}C_p^q
\left(\sqrt{\pi}\right)^{q-D}\Gamma\left(\frac{D-q}{2}\right)A\left(1,\cdots,1;
\frac{D-q}{2}\right)
\end{equation}

\noindent where the Epstein zeta function
$A\left(a_1,a_2,\cdots,a_p;s\right)$ is defined as

\begin{equation}
A\left(a_1,a_2,\cdots,a_p;s\right)=\sum^{\hspace{0.2cm}\infty\hspace{0.2cm}\prime}_{\{n\}=-\infty}
\left(\sum^{p}_{j=1}a_jn_j^2\right)^{-s}\hspace{3cm}
\end{equation}

\noindent where the prime means that the term
$n_1=n_2=\cdots=n_p=0$ has to be excluded.

Using Eqs.(2)-(4) and the Mellin transformation of $\exp(-b\tau)$

\begin{equation}
\int^{\infty}_0 \tau^{s-1}e^{-b\tau}d\tau=b^{-s}\Gamma(s)
\end{equation}

we have

\begin{eqnarray}
& &\varepsilon_p^D\left(L_1,L_2,\cdots,L_p\right)\hspace{3cm}\nonumber\\
&=& L_p\varepsilon_{p-1}^D
\left(L_1,L_2,\cdots,L_{p-1}\right)\hspace{3cm}\nonumber\\
  &+& \frac{\pi^{(D-p)/2}}{2^{D-p+2}}\Gamma\left(-\frac{D-p}{2}\right)
  E_{p-1}\left(\frac{1}{L_1^2},\cdots,\frac{1}{L_{p-1}^2};-\frac{D-p}{2}\right)
  \hspace{3cm}\nonumber\\
 &-& \frac{1}{2^{D-p+1}}L_p^{-\frac{D-p}{2}}\sum^{\infty}_{k=0}\frac{(16\pi)^{-k}}{k!}
 L_p^{-k}\prod^k_{j=1}\left[(D-p+1)^2-(2j-1)^2\right]\hspace{3cm}\nonumber\\
 &\times&\sum^{\infty}_{\{n\}=1}n_p^{(p-D-2-2k)/2}\left(\frac{n_1^2}{L_1^2}+
\cdots+\frac{n_{p-1}^2}{L_{p-1}^2}\right)^{\frac{D-p-2k}{4}}\exp\left[-2\pi
L_pn_p \left(\frac{n_1^2}{L_1^2}+
\cdots+\frac{n_{p-1}^2}{L_{p-1}^2}\right)^{\frac{1}{2}}\right]\hspace{1cm}
\end{eqnarray}

\noindent and

\begin{equation}
\varepsilon_p^D(L)=\frac{L^{p-D}}{2^{D+1}}\pi^{-D/2}\int^{\infty}_0
d\tau\left(\sqrt{\tau}\right)^{D-2}\left\{\bigg[1-\left[\frac{\pi}{\tau}\right]
^{\frac{1}{2}}\bigg]^p-\bigg[\theta_3\left(0,e^{-\tau}\right)-\left[\frac{\pi}{\tau}\right]
^{\frac{1}{2}}\bigg]^p\right\}
\end{equation}

\noindent where the elliptic $\theta$ function

\begin{equation}
\theta_3(0,q)\equiv\sum^{\infty}_{m=-\infty}q^{m^2}\hspace{3cm}
\end{equation}

\noindent when $p=2j+1$ ($j$ is a positive integer), it is
obvious that $\varepsilon_p^D(L)<0$ for any $D$, since
$\theta_3(0,e^{-\tau})>1$ and integrand is always negative
between the integration limits. On the other hand, numerical
calculations show that the energy density is positive for $p$ is
even and $D\leq 6$ in the $p=2j$ case [9]. Noticed that the
terminology "Casimir force" for $L_i$ direction is in proportion
to the derivative of Casimir energy with respective to $L_i$, we
need to study firstly the behaviour of Casimir energy so as to
discuss the nature of Casimir force.We can prove the following
lemmata for the Casimir energy.

{\it Lemma 1} \hspace{0.8cm} For two spacetime dimensions $D_1$
and $D_2$, if $D_2>D_1$, and $\varepsilon_{2j}^{D_1}(L)\leq 0$,
then $\varepsilon_{2j}^{D_2}<0$.

{\it Proof} \hspace{0.8cm} Defining

\begin{equation}
k(\tau)\equiv\left[1-\left(\frac{\pi}{\tau}\right)^{\frac{1}{2}}\right]^{2j}
-\left[\theta_3\left(0,e^{-\tau}\right)-\left(\frac{\pi}{\tau}\right)^{\frac{1}{2}}\right]^{2j}
\end{equation}

\noindent which has only one real root $0<\tau_0<\pi$ for any
$j$. Since $k(\tau)>0$ for $0<\tau<\tau_0$ and $k(\tau)<0$ for
$\tau_0<\tau<\infty$, we have

\[
\varepsilon_{2j}^{D_2}(L) =\frac{L^{p-D_2}}{2^{D_2+1}}\pi^{-D_2/2}
\left(\int^{\tau_0}_0+\int^{\infty}_{\tau_0}\right)d\tau\left(\sqrt{\tau}\right)
^{D_1+(D_2-D_1)-2}k(\tau)
\]

\begin{equation}
<\left(\frac{1}{2L}\sqrt{\frac{\tau_0}{\pi}}\right)^{D_2-D_1}
\varepsilon_{2j}^{D_1}(L)\leq 0
\end{equation}

{\it Lemma 2} \hspace{0.8cm}There exists a particular of spacetime
dimension $D_c=6$, such that for $D\leq D_c,
\varepsilon_2^D(L)>0$ and for $D > D_c, \varepsilon_2^D(L)<0$.

{\it Proof} \hspace{0.8cm}In the $p=2$ case, Eq.(7) can be reduced
to

\begin{equation}
\varepsilon_2^D(L)=\frac{L^{2-D}}{2^{D-1}\pi^{D/2}}\left[2\sqrt{\pi}
\Gamma\left(\frac{D-1}{2}\right)
\zeta\left(\frac{D-1}{2}\right)\beta\left(\frac{D-1}{2}\right)-
\Gamma\left(\frac{D}{2}\right)
\zeta\left(\frac{D}{2}\right)\beta\left(\frac{D}{2}\right)\right]
\hspace{3cm}
\end{equation}

\noindent Where $\zeta(r)$ is the Riemann $\zeta$ function and
$\beta(r)$ is the Dirichlet series

\begin{equation}
\beta(r)\equiv\sum^{\infty}_{j=0}\frac{(-1)^j}{(2j+1)^r}\hspace{3cm}
\end{equation}

\noindent Using these known one-dimensional series, we have
$\varepsilon_2^3(L)=0.04104, \varepsilon_2^4(L)=0.00483,
\varepsilon_2^5(L)=0.00081, \varepsilon_2^6(L)=0.00011$ and
$\varepsilon_2^7(L)=-1.9\times 10^{-5}$ if $L$ is the chosen unit
length. thus, we show Lemma 2 from Lemma 1.

{\it Lemma 3} \hspace{0.8cm}If $j$ is large enough, then
$\varepsilon_{2j}^D(L)<0$.

{\it Proof} \hspace{0.8cm}For $\tau>4\pi>\tau_0$, one can easily
find

\begin{equation}
\left|k(\tau)\right|>\frac{pe^{-\tau}}{2^{2j-2}}
\end{equation}

\noindent From Eq.(7), we have

\begin{eqnarray}
\varepsilon_{2j}^D(L)&<&\frac{L^{2j-D}}{2^{D+1}}\pi^{-\frac{D}{2}}\left(2\pi^j
\tau_0^{\frac{1}{2}}-\int^{\infty}_{64\pi}d\tau\tau^{(p-1)/2}\left|k(\tau)\right|\right)
\nonumber\\
&<&2^D\left(4\sqrt{\pi}L\right)^{2j-D}\left[4^{-2j}\tau_0^{\frac{1}{2}}
-p\left(4\sqrt{\pi}\right)^{-1}e^{-64\pi}\right]
\end{eqnarray}

\noindent then, for $j$ large enough, $\varepsilon_{2j}^D(L)<0$.

{\it Lemma 4} \hspace{0.8cm}If $D$ is large enough, then
$\varepsilon_{2j}^D(L)<0$ for any $j$.

{\it Proof} \hspace{0.8cm}From Eq.(7), we have

\begin{eqnarray}
\varepsilon_{2j}^D(L)&<&\frac{L^{2j-D}}{2^{D+1}}\pi^{-\frac{D}{2}}
\left(\int^{\tau_0}_0+\int^{\infty}_{\tau_0}\right)d\tau\left(\sqrt{\tau}\right)
^{D-2}k(\tau)\nonumber\\
&<&\frac{L^{2j-D}\tau_0^{\frac{D}{2}-1}}{2^{\frac{D}{2}+2}\pi^{\frac{D}{2}}}
\left[\frac{1}{2^{\frac{D}{2}-1}}\int^{\tau_0}_0 d\tau
k(\tau)+\int^{\infty}_{\tau_0} d\tau k(\tau)\right]
\end{eqnarray}

\noindent Since the first term is positive and the second term is
negative in the brackets of Eq.(15), when $D$ is large enough,
then $\varepsilon_{2j}^D(L)<0$ for any $j$.

 {\it Lemma 5} \hspace{0.8cm}The Casimir force of a
$p$-dimensional rectangular cavity with the Dirichlet boundary
conditions can be written as terms of ($p$-1)-dimensional Casimir
energy density and multiseries with exponential factors.

{\it Proof} \hspace{0.8cm}From the recursion relation of the
Casimir energy density Eq.(6), per unit area

\begin{eqnarray}
& &-\frac{\partial \varepsilon_p^D(L_1,\cdots,L_p)}{\partial
L_p}\nonumber\\
=&-&\varepsilon_{p-1}^D(L_1,\cdots,L_{p-1})\nonumber\\
&+&\frac{1}{2^{D-p+2}}
\sum^{\infty}_{k=0}\frac{(16\pi)^{-k}}{k!}(p-D-2k)L_p^{\frac{p-D-2k-2}{2}}
\prod^k_{j=1}\left[(D-p+1)^2-(2j-1)^2\right]\nonumber\\
&\times&\sum^{\infty}_{\{n\}=1}n_p^{\frac{p-D+2+2k}{2}}\left(\frac{n_1^2}{L_1^2}
+\cdots+\frac{n_{p-1}^2}{L_{p-1}^2}\right)^{\frac{D-p-2k}{4}}\exp\left[-2\pi
L_pn_p\left(\frac{n_1^2}{L_1^2}
+\cdots+\frac{n_{p-1}^2}{L_{p-1}^2}\right)^{\frac{1}{2}}\right]\nonumber\\
&-&\frac{1}{2^{D-p}}
\sum^{\infty}_{k=0}\frac{(16\pi)^{1-k}}{k!}L_p^{\frac{p-D-2k}{2}}
\prod^k_{j=1}\left[(D-p+1)^2-(2j-1)^2\right]\nonumber\\
&\times&\sum^{\infty}_{\{n\}=1}n_p^{\frac{p-D+4+2k}{2}}\left(\frac{n_1^2}{L_1^2}
+\cdots+\frac{n_{p-1}^2}{L_{p-1}^2}\right)^{\frac{D-p-2k+2}{4}}\exp\left[-2\pi
L_pn_p\left(\frac{n_1^2}{L_1^2}
+\cdots+\frac{n_{p-1}^2}{L_{p-1}^2}\right)^{\frac{1}{2}}\right]
\end{eqnarray}

\noindent This completes the proof of Lemma 5.

Since the second and the third terms are obviously negative in the
right hand of Eq.(16), so that the pressure is always
negative(directed inwards) if the first term is also negative.
From Lemmata 1-4, we show that there exist a critical value of the
spacetime dimension $D_c$ for which $\varepsilon_{2j}^D>0$ if
$D\leq D_c$ and $\varepsilon_{2j}^D<0$ if $D> D_c$. However,
$2j+1>D_c$ if $j>14$. In this case there is no critical $D_c$
since all $\varepsilon_{2j}^D$(for $D=2j+1, 2j+2, \cdots$) are
negative. We summarize the above in the following.

{\it Theorem}\hspace{0.8cm}The Casimir effect gives rise to an
attractive force between the configuration boundaries that
confine the massless scalar field for $p$-odd hypercube with the
Dirichlet boundary conditions and $D\leq D_c, p\leq 29$.

Furthermore,  we show the dependence of the critical value $D_c$
on $p$ in Table 1 using numerical calculation for possible
physical application.

\vspace{1cm} \textbf{ Table 1}  The critical value $D_c$ for
massless scalar fields satisfying Dirichlet boundary conditions
inside a hypercube with $p$-odd unit sides in a $D$-dimensional
spacetime.

------------------------------------------------------------------------------------------------

$p$\hspace{0.5cm}3\hspace{0.5cm}5\hspace{0.5cm}7\hspace{0.5cm}9
\hspace{0.5cm}11\hspace{0.5cm}13\hspace{0.5cm}15\hspace{0.5cm}17
\hspace{0.5cm}19\hspace{0.5cm}21\hspace{0.5cm}23\hspace{0.5cm}25
\hspace{0.5cm}27\hspace{0.5cm}29

-------------------------------------------------------------------------------------------------

$D_c$\hspace{0.2cm}7\hspace{0.5cm}9\hspace{0.4cm}11\hspace{0.3cm}12
\hspace{0.4cm}14\hspace{0.5cm}16\hspace{0.5cm}17\hspace{0.5cm}19
\hspace{0.5cm}21\hspace{0.5cm}23\hspace{0.5cm}24\hspace{0.5cm}26
\hspace{0.5cm}28\hspace{0.5cm}30

--------------------------------------------------------------------------------------------------

Finally, we shall give a brief discussion on our result. In spite
of an impressive literature on the Casimir effect[10], the query
whether its attractive or repulsive character changes by going to
higher dimensions had never been elucidated for Dirichlet
boundary conditions. We analytically show that Casimir force of
$p$-odd hypercube is attractive, in contrast with the result of
Ref.[11]. Our result is consistent with numerical calculation[9].
It may be worth emphasizing that Epstein zeta function is the
fundamental zeta function associated with higher dimensions.

\bigskip
\noindent\textsc{acknowledgments}

\noindent This work is supported by the National Natural Science
Foundation of China under Grant No. 19875016 , the Doctorial Fund
from the Education Ministry of China under Grant No. 1999025110
and the Foundation for the Development of Science and Technology
of Shanghai under Grant No.01JC1435.

\end{document}